\documentclass{aa}

\usepackage[varg]{txfonts}

\usepackage{multirow}
\usepackage{graphicx}

\def\ms{\mbox{$M_{\rm s}$}}
   
\def\mstar {$M_{\star}$}

\def\msun {$M_{\sun}$}
\def\Q {$Q_{\rm LSS}$}
\def\etak {$\eta_{\rm k,LSS}$}
\def\dfive {$d_{\rm 5N}$}

\usepackage[colorlinks, citecolor={blue}]{hyperref}
\bibpunct[; ]{(}{)}{;}{a}{}{;}

\begin{document}

\title{The less significant role of large-scale environment than optical AGN in nearby, isolated elliptical galaxies}

\titlerunning{The large-scale environment of isolated elliptical galaxies}

\author{I.~Lacerna\inst{\ref{ucn},\ref{mas},\ref{ia-puc}} \and 
M.~Argudo-Fern\'andez\inst{\ref{antof},\ref{cassaca}} \and 
S. Duarte Puertas\inst{\ref{iaa}}
}

\institute{
Instituto de Astronom\'ia, Universidad Cat\'olica del Norte, Av. Angamos 0610, Antofagasta, Chile\\
\email{ivan.lacerna@ucn.cl}  \label{ucn}
\and
Instituto Milenio de Astrof\'isica, Av. Vicu\~na Mackenna 4860, Macul, Santiago, Chile
 \label{mas}
\and
Instituto de Astrof\'isica, Pontificia Universidad Cat\'olica de Chile, Av. V.~Mackenna 4860, Santiago, Chile 
\label{ia-puc} 
\and 
Centro de Astronom\'ia (CITEVA), Universidad de Antofagasta, Avenida Angamos 601 Antofagasta, Chile
\label{antof}
\and 
Chinese Academy of Sciences South America Center for Astronomy, China-Chile Joint Center for Astronomy, Camino El Observatorio, 1515, Las Condes, Santiago, Chile
\label{cassaca}
\and
Instituto de Astrof\'isica de Andaluc\'ia IAA – CSIC, Glorieta de la Astronom\'ia s/n, 18008 Granada, Spain \label{iaa}
}

\abstract
%Context
{
The formation and evolution of elliptical galaxies in low-density
environments are less understood than classical elliptical galaxies in high-density environments. 
Isolated galaxies are defined as galaxies without massive neighbors within scales of galaxy groups. 
The effect of the environment at several Mpc scales on their properties has been barely explored.
Here we study the role of the large-scale environment in 573 isolated elliptical galaxies out to $z = 0.08$. 
}
%Aims
{
We aim to explore whether the large-scale environment affects
some of their physical properties.
}
%Methods
{
We use three environmental estimators of the large-scale structure within a projected radius of 5 Mpc around isolated galaxies: the tidal strength parameter,  the projected density 
\etak, and the distance to the fifth nearest neighbor galaxy. 
We study isolated galaxies regarding stellar mass, integrated optical $g-i$ color, specific star formation rate (sSFR), and emission lines. 
}
%Results
{
We find 80\% of galaxies at lower densities correspond to `red and dead' elliptical galaxies. 
Blue and red galaxies do not tend to be located in different environments according to \etak. Almost all the isolated ellipticals in the densest large-scale environments are red or quenched, where a third of them are low-mass galaxies. 
The percentage of isolated elliptical galaxies located in the AGN region of the %\citet[][BPT]{BPT1981}
Baldwin et al. (BPT) diagram is 64\%.
We have identified 33 blue, star-forming isolated ellipticals 
using both color and sSFR.
Half of them are star-forming nuclei in the BPT diagram, 
which is 5\% of the galaxies in this diagram.
}
%Conclusions
{
The large-scale environment is not playing the primary role to determine the color or sSFR of isolated elliptical galaxies. 
The large-scale environment seems to be negligible from a stellar mass scale around $10^{10.6}$ \msun, 
probably 
because of the dominant presence of AGN at 
higher masses. For lower
masses, the processes of cooling and infall of gas from large scales are very inefficient in ellipticals. AGN might also be an essential ingredient to keep most of the low-mass isolated elliptical galaxies quenched.
}

\keywords{galaxies: active  -
galaxies: E and lenticular, cD -   
galaxies: fundamental parameters - galaxies: photometry - 
galaxies: star formation
}

\maketitle

%============================================================
\section{Introduction} 
\label{S1}

Early-type galaxies, and in particular elliptical (E) galaxies, have been described in general as quiescent objects in high-density environments with regular and smooth structures as the result of major and minor mergers of disk galaxies, after which the gas would have been depleted, and star formation quenched \citep[e.g.][]{Oemler1974, Dressler1980,  Hernquist1993, Kauffmann1996, Tutukov+2007, Schawinski+2014}.
However, detailed observational studies in recent years have shown that early-type galaxies present more complex structures and more variations
in their properties than previously thought \citep{BlantonMoustakas2009,Suh+2010,Shapiro+2010,Young+2014}.

Although early-type galaxies are mostly red or passive objects, there is also a fraction of blue or star-forming objects that increase as the mass is smaller and the environment is less dense 
\citep[e.g.,][]{Kuntschner+2002,Bernardi+2006,Lee+2006,Schawinski+2009, Kannappan+2009, Thomas+2010, McIntosh+2014, Schawinski+2014, GZ2015, Vulcani+2015, Lacerna+2016, SpectorBrosch2017}.
However, the formation and evolution of early-type galaxies in low-density environments are less understood than classical early-types in high-density environments \citep{Kuntschner+2002,Rosito+2018}.

Mechanisms that regulate the physical properties of massive E galaxies seem to be concentrated on scales inside the virial radius of their host dark matter haloes. In this case,
the processes involved in the morphological transformation of
E galaxies are those that dominate in their quenching of star formation and the depletion of their cold gas reservoir.
New episodes of cold gas inflow are very unlikely to occur within the high-density environment of clusters or for isolated galaxies living in massive haloes, 
thus, these E galaxies are expected to remain `red and dead.'

The effect of the environment around massive galaxies is 
less strong than mass to define their physical properties
\citep[e.g.,][]{Alpaslan+2015}
but seems to play some role around less massive galaxies \citep{Peng+2010, Bluck+2014}. 
\citet[][hereafter L16]{Lacerna+2016}
find that the correlations of colors, specific star formation rate (sSFR), and size with the stellar mass of nearby isolated E galaxies are similar to those ellipticals in a high-density environment, the Coma supercluster. For instance, all E galaxies more
massive than $10^{11}$ \msun\ are quiescent. 
However, at smaller stellar masses, a fraction of the isolated Es deviate systematically toward the blue cloud, whereas a few Es pass to be moderately blue in Coma.
There are
four isolated Es with stellar masses between 1 $\times$ $10^{10}$ and 4 $\times$ $10^{10}$ \msun\
that are blue and star-forming (SF) galaxies at the same time. These blue
SF isolated Es are also the youngest galaxies with light-weighted stellar ages $\lesssim$ 1 Gyr, which could indicate recent processes of star formation in them. Therefore, in some cases, 
the isolated environment seems to propitiate the rejuvenation or a late
formation of E galaxies.

Most of the definitions of isolated galaxies are based on the lack of neighbor galaxies with comparable brightness or mass within some distance that usually is not larger than 1 Mpc. Therefore, isolation criteria are given by 
the local environment at scales slightly beyond the virial radius of big galaxy groups or small galaxy clusters, but they do not consider the large-scale structures around isolated galaxies. 
In this context, isolated galaxies can be residing in very different large-scale structures such as voids, void walls, filaments, or in the outskirts of galaxy clusters \citep{Argudo-Fernandez+2015}. 
Therefore, the large-scale environment might supply the gas for the rejuvenation scenario in some isolated E galaxies, or, on the other hand, the cosmic web might remove the gas reservoir by ram pressure in some isolated Es hosted by low-mass haloes \citep[e.g.`cosmic web stripping',][]{Benitez-Llambay+2013} among other scenarios.

This paper aims to explore whether the large-scale environment affects the integrated photometric and spectral properties of nearby isolated galaxies with E morphologies as a function of total stellar mass. 
We use three environmental estimators of the large-scale structure within a projected radius of 5 Mpc,  
which is a scale that is typically outside of filament spines \citep{Kuutma+2017}, around isolated galaxies.

The outline of the paper is as follows. 
Section \ref{S2} presents the descriptions of two catalogs of isolated galaxies used in this paper. 
In this section, we also describe the morphological classification of E galaxies, and their physical properties used along the paper. Besides that, here we define the large-scale environmental parameters.
We show both the color-stellar mass diagram and the sSFR-stellar mass diagram as a function of the large-scale environment of isolated E galaxies in Sect. \ref{S3results}.
Section \ref{S4bpt} presents the \citet[][BPT]{BPT1981} %BPT 
diagrams as a function of stellar mass and large-scale environment.
The implications of our results are discussed in Sect. \ref{Sdiscussion} and the conclusions are given in Sect. \ref{Sconclusions}.

Throughout this paper the reduced Hubble constant, $h$, is defined as $H_0 = 70$ $h$ km s$^{-1}$ Mpc$^{-1}$ with the following dependencies: stellar mass in $h^{-2}~\rm M_{\sun}$ and physical scale in $h^{-1}~\rm{ Mpc}$.

%============================================================
\section{Data} 
\label{S2}
%============================================================   

Because the number of very isolated galaxies is low, especially for galaxies of E morphologies, we join two catalogs of isolated galaxies in our analysis; the UNAM-KIAS and SIG samples, which are described in  
Sects. \ref{sampleUNAM-KIAS} and \ref{sampleSIG}, respectively. We present our morphological selection of E galaxies in Sect. \ref{sec_morpho}. 
We present some physical properties that we use to characterize our samples of isolated E galaxies in Sect. \ref{sec_prop}.
In Sect. \ref{sec_env}, we define the large-scale environmental estimators used in this work. 

\subsection{The UNAM-KIAS sample}
\label{sampleUNAM-KIAS}

One sample of isolated galaxies in this study comes from the UNAM-KIAS catalog of \cite{Hernandez-Toledo+2010}. That paper and also L16 give more details of the sample. Here we briefly refer to present this sample and the selection criteria.

Galaxies with extinction-corrected apparent 
Petrosian $r$-band magnitudes in the range $14.5 \leq r_{\rm Pet} < 17.6$ were selected from the 
SDSS Data Release 5 \citep{DR5+2007}.
\cite{Hernandez-Toledo+2010} also used data from the New York University Value-Added Galaxy Catalog 
\cite[NYU-VAGC;][]{Blanton+2005} 
based on SDSS DR4 \citep{DR4+2006}.
The survey region used covers 4464 deg$^2$, containing 312\,338 galaxies. \cite{Hernandez-Toledo+2010}
searched in the literature and borrowed redshifts of the bright galaxies without SDSS spectra to increase the spectroscopic completeness. The final dataset consists of 317\,533 galaxies with known redshift and SDSS photometry. 

Three parameters specify the isolation criteria.
The first is the extinction-corrected Petrosian $r$-band apparent magnitude 
difference between a candidate galaxy and any neighboring galaxy, $\Delta m_r$. The second is the projected separation to the neighbor 
across the line of sight, $\Delta d$. The third parameter is the radial velocity difference, $\Delta V$. 
Assume a galaxy $i$ has a magnitude $m_{r,i}$ and $i$-band Petrosian radius $R_i$. It is regarded as isolated with respect to potential perturbers if the separation $\Delta d$ between this galaxy and a neighboring galaxy $j$ with magnitude $m_{r,j}$ and radius $R_j$ satisfies the conditions
\begin{equation}
\Delta d \ge 100\times R_j  
\label{eq1_UNAM-KIAS}
\end{equation}
\begin{equation}
\textrm{or }  \rm 
\Delta V \ge 1000 \textrm{ km s}^{-1}. 
\label{eq2_UNAM-KIAS}
\end{equation}
Here $R_{j}$ is the seeing-corrected Petrosian radius of galaxy $j$, measured in $i$-band using elliptical annuli to consider flattening or inclination of galaxies \citep{Choi+2007}. 
In cases when a galaxy $i$ has close neighbors (i.e., it does
not satisfy Eqs. \ref{eq1_UNAM-KIAS} and \ref{eq2_UNAM-KIAS}), this galaxy can be considered as isolated if
\begin{equation}
m_{r,j} \ge m_{r,i} + \Delta m_r, 
\label{eq5_UNAM-KIAS}
\end{equation}
for all neighboring galaxies. 
\cite{Hernandez-Toledo+2010} chose $\Delta m_r = 2.5$ mag. 
Only galaxies brighter than $m_r = 15.2$ mag were used to select isolated galaxies to 
consider this magnitude difference.
This condition allows an isolated galaxy to have fainter neighbors but rejects significant perturbers.
Using these criteria, they found a total of 1520 isolated galaxies.

\subsection{The SIG sample}
\label{sampleSIG}

The other sample of isolated galaxies used in this study is the SDSS-based catalog of Isolated Galaxies compiled by \citet{Argudo-Fernandez+2015}, hereafter SIG. The SIG is composed of 3702 
galaxies selected from the SDSS-DR10 \citep{Ahn+2014}. The galaxies were selected in a volume-limited sample with SDSS $r$-band model magnitude $11 \leq m_{r} \leq 15.7$ and redshift range $0.005 \leq z \leq 0.080$, and are isolated with no neighbors within projected separation and line-of-sight velocity difference as follows:
\begin{equation}
\Delta d \le 1\,\rm{Mpc}
\label{eq1_SIG}\end{equation}
\begin{equation}
\textrm{and }|\Delta V| \le 500 \textrm{ km s}^{-1}.
\label{eq2_SIG}\end{equation}

The isolation criteria used in the SIG are slightly different from the criteria in the UNAM-KIAS sample. The volume of isolation in the SIG is fixed at 1\,Mpc projected distance in the sky, which corresponds to a crossing time of about $t_{cc}\sim 5$\,Gyr \citep[see][for more details]{Argudo-Fernandez+2015}. 
The upper magnitude limit for galaxies in the SIG at $m_{r} \leq 15.7$\,mag ensures that there are no neighbors, in the defined isolation volume, at least 2 orders of magnitude fainter within the range of spectroscopic completeness of the SDSS main galaxy sample, at $m_{r,\rm{Petrosian}}~<~17.77$\,mag. 
This limit is roughly similar to the magnitude limit of 15.2 mag of the UNAM-KIAS sample. There is an overlap with the same sky area of almost 4500 deg$^2$. 
These features finally translate
in a good agreement between the two criteria, where 43.5\% of the galaxies in the UNAM-KIAS are found in the SIG.

\subsection{Morphology}
\label{sec_morpho}

The isolated galaxies from the UNAM-KIAS catalog were classified morphologically after some basic image processing. 
The processing included surface brightness profiles and the corresponding geometric profiles 
(ellipticity $\epsilon$, Position Angle PA and $A_{4}/B_{4}$ coefficients of the Fourier series expansions of deviations of a pure ellipse) 
from the $r$-band images to provide additional evidence of boxy or disky character 
and other structural details. In particular, a galaxy was judged to be an 
E if the $A_{4}$ parameter showed: 1) no significant  boxy ($A_{4} < 0$) or disky ($A_{4} > 0$) trend in the outer parts, or 2) a 
generally boxy ($A_{4} < 0$) character in the outer parts. The isolated galaxies were inspected for the presence or absence of 3) a linear component in the surface brightness-radius diagram. Morphologies were assigned according to a numerical code following the HyperLeda\footnote{http://leda.univ-lyon1.fr/} 
database convention; in particular for early-type galaxies, the following $T$ morphological parameters \citep{Buta+1994} are applied: $-5$ for E, $-3$ for E-S0, $-2$ for S0s, and 0 for S0a types. In the UNAM-KIAS sample, 
92 isolated E galaxies satisfy
$T \le -4$, which is $\approx 6\%$ of the sample. 
For a fair comparison between the two samples of isolated galaxies, hereafter we consider 89 isolated E galaxies within the redshift range
0.0041 $\le z \le$ 0.08.

There is not an available morphological classification for SIG galaxies, in this case we therefore used the morphological classification for SDSS galaxies provided by \citet[HC11,][]{HC+2011}, Galaxy Zoo 1 \citep[GZ1,][]{GZC+2011}, and Galaxy Zoo 2 \citep[GZ2,][]{Willett+2013}. 
In HC11, they considered galaxies with spectroscopic redshift $z<0.25$ and computed the probability distribution of morphologies for galaxies brighter than $m_r < 16$ mag using a Bayesian automated classification. Each galaxy is associated with six values of probabilities (early types, late types, elliptical, lenticular, Sab, and both Scd and Im), where the probability for classes elliptical, lenticular, Sab, and both Scd and Im sums to unity. 
There are 3467 SIG galaxies with morphological classification in HC11. 
Here we use the probability to be an E galaxy, $probEll$.
On the other hand, GZ1 and GZ2 are citizen science projects for visual morphological classification of galaxies. In GZ1, galaxies are classified into three categories: $SPIRAL$, $ELLIPTICAL$, and $UNCERTAIN$. Here $ELLIPTICAL$ corresponds to galaxies with early-type morphologies. To classify the galaxies morphologically with GZ2, we have converted the morphological types given by
\cite{Willett+2013}
to values of numerical $T$-type (from -5 to 12, from elliptical
to irregular, respectively). 
There are 3514 and 2726 SIG galaxies with morphological classification in GZ1 and GZ2, respectively.

We check that for the E galaxies from the UNAM-KIAS sample, the median value of $probEll$ is 0.76, 90\% of them are $ELLIPTICAL$ in GZ1, and 100\% of them have $T = -5$ in GZ2.

We consider a SIG galaxy as an E galaxy according to the condition
\begin{equation}
probEll \geq 0.85 . 
\label{SIG_cond1} \end{equation}
This is a robust condition to select E galaxies after visual inspection. 
From the results using E galaxies in the UNAM-KIAS sample, we find that the previous condition can be slightly relaxed as long as we include morphological classifications from either GZ1 or GZ2. Therefore, a SIG galaxy is also considered as E with the following conditions
\begin{gather}
\label{SIG_cond2a} 
ELLIPTICAL = 1 \\
\textrm{and }  probEll > 0.7 
\label{SIG_cond2b} \end{gather}
where we join the classifications from HC11 and GZ1,
or the conditions
\begin{gather}
\label{SIG_cond3a}
T=-5 \\
\textrm{and }  probEll > 0.7  
\label{SIG_cond3b} \end{gather}
where we join the classifications from HC11 and GZ2.
Finally, we find that taking into account GZ1 and GZ2 samples together we obtain reliable classifications of E galaxies. Thus, we also use 
the conditions
\begin{gather}
\label{SIG_cond4a}
ELLIPTICAL = 1 \\
\textrm{and } T = -5.
\label{SIG_cond4b} \end{gather}
As a summary, SIG galaxies are classified as E galaxies using Eq. (\ref{SIG_cond1}), or Eqs. (\ref{SIG_cond2a})--(\ref{SIG_cond2b}), or Eqs. 
(\ref{SIG_cond3a})--(\ref{SIG_cond3b}), or Eqs. (\ref{SIG_cond4a})--(\ref{SIG_cond4b}).

There are 504 ($\sim 13.6\%$) SIG galaxies classified as ellipticals using the conditions described above. We find 20 E galaxies from the SIG catalog 
are in the sample of ellipticals in the UNAM-KIAS catalog. Therefore, our sample is made of 573 
distinct isolated E galaxies from both UNAM-KIAS and SIG samples.

\subsection{Integrated physical properties}
\label{sec_prop}

Many public databases provide
physical parameters for 
galaxies in the SDSS as stellar masses, sSFRs, emission lines, and colors \citep[e.g.,][]{Kauffmann+2003,Brinchmann+2004,Blanton+2005,Salim+2007,Maraston+2013,DuartePuertas+2017}. However, not all the galaxies in our samples have available information. To avoid reducing the sample and to be consistent along the present work, we apply the same methodology to obtain colors and sSFR, and to estimate the stellar mass for our isolated E galaxies.

Color measurements were taken from the SDSS database with extinction corrected $modelMag$ magnitudes in DR12 \citep[][]{DR12+2015}. 
This magnitude (\verb|dered| parameter in CasJobs\footnote{http://casjobs.sdss.org/CasJobs/}) is defined as the better of two magnitude fits: a pure de Vaucouleurs profile and a pure exponential profile.

Stellar masses were estimated by fitting the spectral energy distribution (SED) to the five SDSS photometric bands ($u, g, r, i, z$) using the routine \verb|kcorrect| \citep{Blanton+2007}. In their latest version, SEDs are based on unaltered stellar evolution synthesis results from Bruzual-Charlot models \citep{BC2003}, therefore for each fit galaxy \verb|kcorrect| provides an estimate of the stellar mass. 
We note that Bruzual-Charlot models do not contain extreme horizontal branch stars that might provide more of the UV excess
seen in E galaxies \citep{GreggioRenzini1990,Brown+2000}.
The lack of these stars
may cause an underestimation of the stellar mass for more massive E galaxies. We checked the difference is not higher than 0.11 dex on average for a sample of E galaxies than the mass-to-light ratio of \citet{Bell+2003} that includes a correction to Petrosian magnitudes for galaxies with this morphology to estimate their stellar mass. This difference is within the range of the systematic error of the mass-to-light ratio method (0.1--0.15 dex). Therefore, our stellar mass estimation is reliable within the typical uncertainties.  

The sSFR has been obtained from the Max Plank Institute for Astrophysics and the Johns Hopkins University \citep[MPA-JHU\footnote{Available at \texttt{http://www.mpa-garching.mpg.de/SDSS/DR7/}};][]{Kauffmann+2003,Tremonti+2004,Salim+2007} added value catalog \citep{Brinchmann+2004} that uses a spectrophotometric synthesis fitting model.

From the same MPA-JHU catalog, we use the emission line fluxes 
of H$\alpha$, H$\beta$, [OIII]$\lambda$5007, and [NII]$\lambda$6584
to study the BPT diagram \citep{BPT1981, VeilleuxOsterbrock1987, Kewley+2001,Kauffmann+2003}.
In particular, we consider those lines where the flux is positive and the signal-to-noise ratio (SNR) is $\ge$ 3, i.e., strong line galaxies (SLGs) according to \cite{CidFernandes+2010}. 
This is the typical cut used for SDSS galaxies 
\citep[e.g.,][]{Kauffmann+2003,Brinchmann+2004,Li+2006}.
Below this limit, it is difficult to make a robust interpretation of the BPT diagram.
It is also necessary to have a reliable measurement of the H$\alpha$ emission-line flux to study the blue SF galaxies adequately.
However, to study a possible bias that can be introduced by the SLGs definition, we also consider weak emission-line galaxies (WLGs) based on definitions from \cite{CidFernandes+2010} in Appendix \ref{Ap_wlg}. Our results are consistent when we consider only SLGs and both SLGs and WLGs. Thus we will focus on SLGs in the main body of the paper.

\subsection{Large-scale environmental parameters}
\label{sec_env}

\citet{Argudo-Fernandez+2015} provided quantification of the large-scale environment for galaxies in the SIG. They use two parameters: the tidal strength parameter \Q\,(Eq.~\ref{Eq:Q3}) and the projected density \etak\,(Eq.~\ref{Eq:etak3}). The large-scale 
structure (LSS) is characterized considering all the neighbors around each isolated galaxy in a volume of 5\,Mpc projected distance radius within a line-of-sight velocity difference of $|\Delta\,V|~\leq~500$\,km\,s$^{-1}$.

For each SIG galaxy, the \Q\, exerted by all the $i$ galaxies of its LSS, located at a projected distance $d_{i}$, is defined as:

\begin{equation} \label{Eq:Q3}
Q_{\rm LSS} \equiv {\rm log} \left(\sum_i {\frac{M_{i}}{M_{\rm SIG}}} \left(\frac{D_{\rm SIG}}{d_i}\right)^3\right),
\end{equation} 
where $M$ is the stellar mass and $D_{\rm SIG}$ is the estimated diameter of the SIG galaxy\footnote{The parameter $D_{\rm SIG} = 2\alpha r_{90}$ is the estimated diameter of the SIG galaxy, where $r_{90}$, the Petrosian radius containing 90\% of the total flux of the galaxy in the $r$-band, is scaled by a factor $\alpha=1.43$ to recover the $D_{25}$ \citep[see][for details]{Argudo-Fernandez+2013}}.

The projected density to the $k^{\rm th}$ nearest neighbor is defined as: 
\begin{equation} \label{Eq:etak3}
\eta_{k, \rm LSS} \equiv \log \left(\frac{k - 1}{\rm{Vol}(d_k)}\right) = \log \left(\frac{3(k - 1)}{4\pi\,d_k^3}\right).
\end{equation} 
\citet{Argudo-Fernandez+2015}  consider $k$ equal to 5, 
therefore $d_k$\,=\,\dfive{} 
is the projected distance to the $5^{\rm th}$ nearest neighbor 
(or less if there were not enough neighbors in the field out to 5 Mpc).

We adopt these two parameters and add the \dfive\ as a third environmental parameter to investigate the large-scale environment of isolated ellipticals. 
We follow the same methodology to quantify the environment of the isolated E galaxies in the L16 sample.

By definition, the tidal strength is a mass-dependent parameter, 
whereas the projected density is a mass-independent parameter. For this reason, these two parameters are complemented very well to characterize the environment around galaxies. The values of \Q\, and \etak\, are both small if the galaxy is located in a low-density environment, and high if the galaxy is located in a dense environment. If a galaxy is part of a small, isolated compact group, for instance, the projected density will have a modest value, 
but the tidal strength 
will be high. On the opposite, a galaxy with a high projected density 
but a small tidal strength could be a galaxy in a large group or small cluster with large distances between members and 
composed of small galaxies. The addition of the third parameter \dfive, even if it is correlated to \etak, allows us to have a more comprehensible sense of the distance at which the \etak\, is calculated. 

\citet{Argudo-Fernandez+2016} found that the \Q\, parameter is correlated very well with the location of isolated systems with respect to their large-scale environment. The lowest values of \Q\, correspond to isolated galaxies mainly located in voids, and gradually increasing values correspond to isolated galaxies closer to the outskirts of more denser structures. In this regard,  isolated galaxies mainly belong to the outer parts of filaments, walls, and clusters, and only about the 30\% of the isolated galaxies are void galaxies 
\citep[see Appendix \ref{Ap_lss} and also][]{Argudo-Fernandez+2015}.

%%%Fig1 
\begin{figure}[h!]
\includegraphics[width=\linewidth]{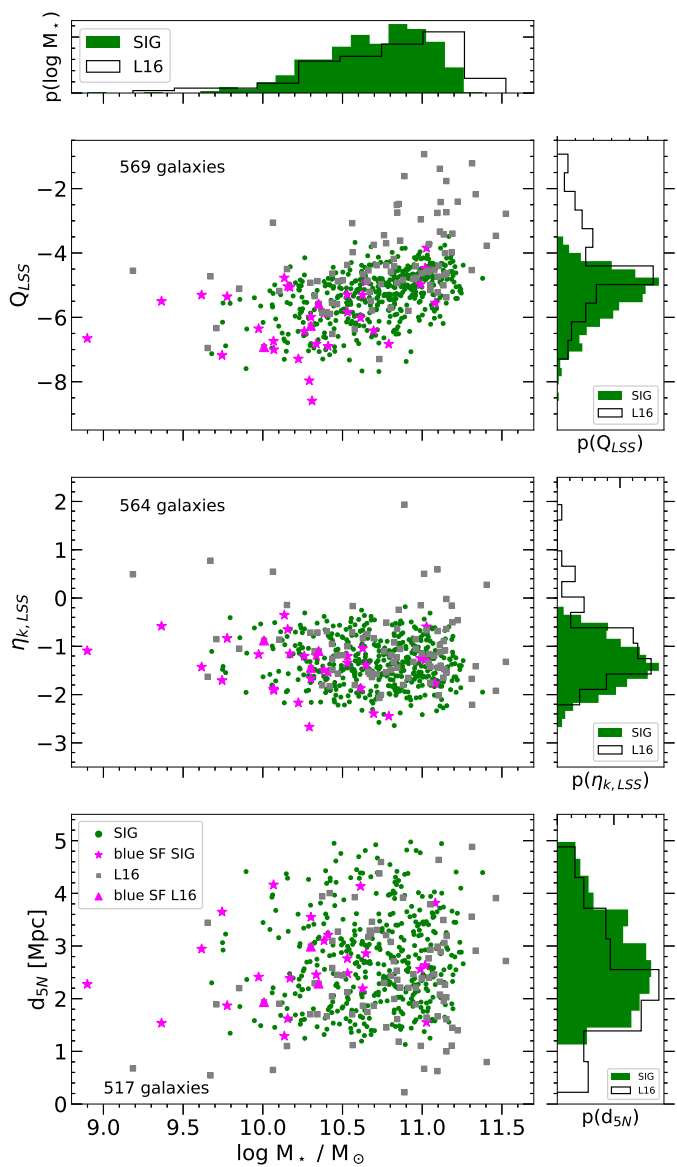}
\caption{Main panels (from top to bottom): as a function of stellar mass, tidal strength estimation of the LSS (\Q), the projected density estimation of the LSS (\etak{}), and the distance to the fifth nearest neighbor galaxy out to 5 Mpc (\dfive). Green dots and grey squares correspond to isolated E galaxies from the SIG and L16 samples, respectively (the total number of galaxies is indicated in each panel). The blue, star-forming isolated ellipticals of SIG sample and L16 sample are shown as magenta triangles and magenta stars, respectively. Top-left panel: normalized density distribution of stellar mass for SIG galaxies (green solid histogram) and L16 galaxies (black open histogram) with estimated \Q{} values.
Right panels (from top to bottom): normalized density distributions of \Q{}, \etak{}, and \dfive{} environmental parameters for SIG galaxies (green solid histogram) and L16 galaxies (black open histogram).
The integral of each histogram sums to unity.
}
\label{fig:lss_Ms}
\end{figure} 

Figure \ref{fig:lss_Ms} shows the environmental parameters of the LSS 
\Q{}, \etak{}, and \dfive\ as a function of stellar mass for the samples of isolated E galaxies described in this section.
\Q{} slightly increases with the mass. This trend is stronger for the most massive galaxies that correspond to the L16 sample (grey squares). In the middle-left panel, we can see that \etak{} is an independent parameter with the stellar mass.
Similarly, \dfive\ is independent of the stellar mass as can be seen in the bottom-left panel.

The top-left panel of Fig. \ref{fig:lss_Ms} shows the histograms with the normalized density distributions of stellar mass for SIG and L16 isolated E galaxies. 
These normalized density distributions were obtained using the Knuth method for estimating the bin width implemented on \textsc{astroML}\footnote{http://www.astroml.org/} \citep{astroML,astroMLText}.
Both mass distributions are relatively similar, except in the high-mass end where the distribution is skewed for more massive galaxies in the L16 sample. The latter can be explained as a combination of the slightly brighter magnitude limit in the UNAM-KIAS sample and 
that it is still possible to select isolated galaxies with much fainter neighbors within a distance of 1 Mpc in this catalog, in contrast to the SIG sample. Therefore, it is more likely to select massive isolated galaxies using the criteria of the UNAM-KIAS catalog 
than the criteria of the SIG catalog.
Right panels 
show histograms with the normalized density distributions of \Q{}, \etak{}, and \dfive{} environmental parameters (from top to bottom) for 
both samples of isolated E galaxies. There is a tail of L16 galaxies with \Q{} > -3.5 that is not present for SIG galaxies. They are in general massive galaxies (66\% of them with \mstar{} $\gtrsim$ 10$^{11}$ \msun). In addition, there is a tail of L16 galaxies with \etak{} > 0, which mostly correspond to galaxies with values of \dfive\ < 1 Mpc. By definition, there are no SIG galaxies with galaxy neighbors to distances shorter than 1 Mpc (see Eq. \ref{eq1_SIG}). As mentioned in Sect. \ref{sampleSIG}, there are some differences in the isolation criteria between SIG and L16 samples, which is reflected in the tails of the distributions of the stellar mass and environmental parameters, but there is an overall agreement in these distributions between both samples. Recall we do not pretend to compare these samples, we aim to join them to increase the total number of isolated E galaxies. Nevertheless, in some cases, we will present some results for both samples separately below.

%========================
\section{Color- and sSFR-stellar mass diagrams as a function of the large-scale environment} 
\label{S3results}
%========================

\begin{figure}
\includegraphics[width=9.3cm]{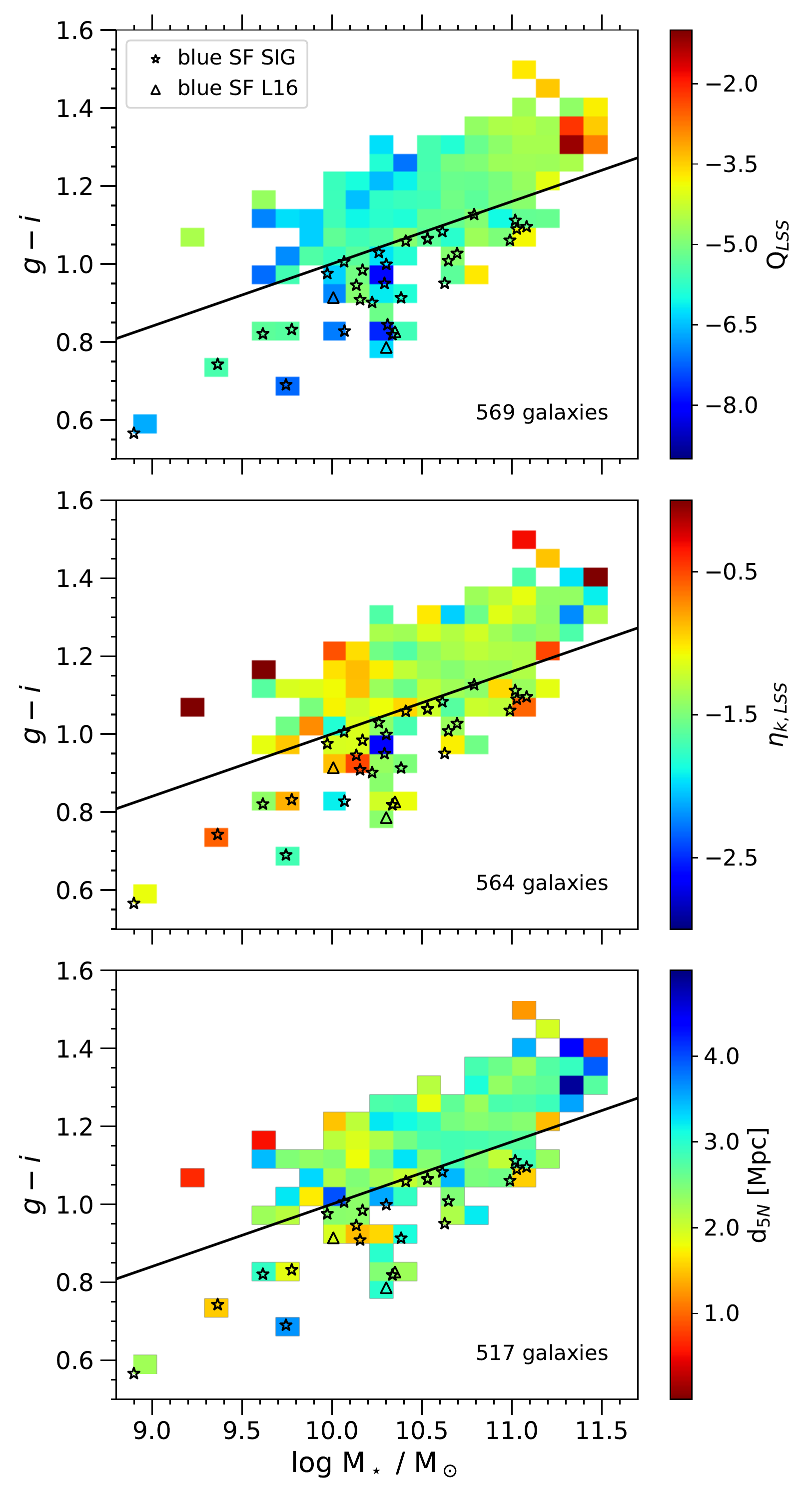}
\caption{
As a function of stellar mass, $g - i$ color of isolated E galaxies.
The black line shows Eq. (\ref{eq_gi}) to separate red and blue galaxies. 
The open black stars and open black triangles show the blue SF isolated ellipticals from the SIG and L16 samples, respectively.
The color bars correspond to the \Q{} parameter (top panel), \etak{} parameter (middle panel), and the distance to the fifth nearest neighbor galaxy out to 5 Mpc (bottom panel). 
The total number of galaxies is indicated in each panel.
}
\label{fig:lss_colorMs}
\end{figure}

Figure \ref{fig:lss_colorMs} shows the $ g - i$ color of isolated E galaxies as a function of stellar mass. We use the relation found by \citet{Lacerna+2014} to separate red and blue galaxies, specifically
\begin{eqnarray} 
(g - i)  = 0.16[\textrm{log($M_{\star}$)}-10.31]+1.05  \textrm{,} 
\label{eq_gi}
\end{eqnarray}
%\\
where the stellar mass is in units of \msun. This relation is shown as a black solid line in the figure. Most of the isolated E galaxies are red galaxies (88\%), 
that is, they are located above the black line.

%%%Fig2 

\begin{figure}
\includegraphics[width=9.3cm]{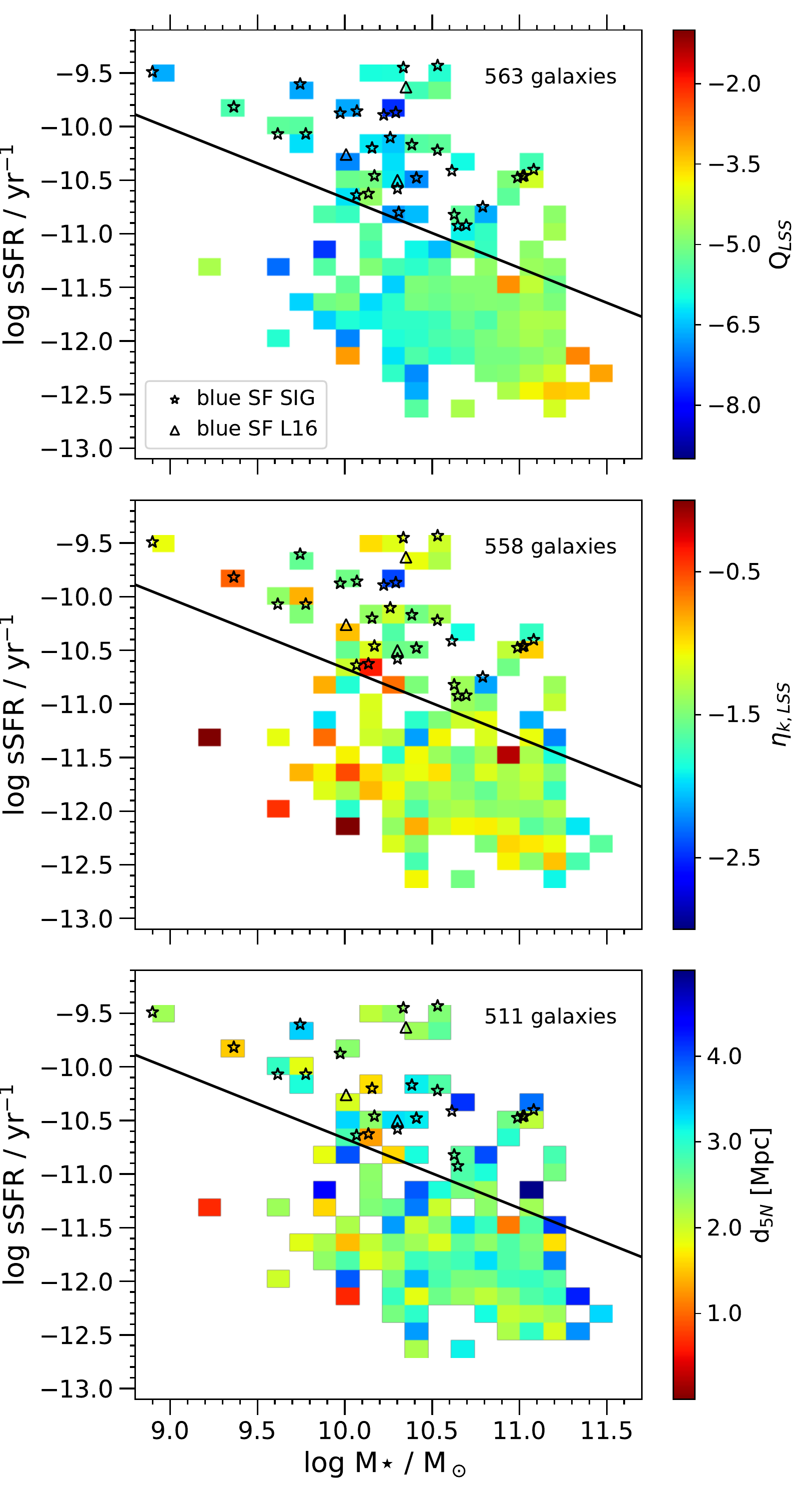}
\caption{
Specific star formation rate (sSFR) as a function of the stellar mass of isolated E galaxies.
The black line shows Eq. (\ref{eq_sSFR}) to separate SF and quenched galaxies. 
The open black stars and open black triangles show the blue SF isolated ellipticals from the SIG and L16 samples, respectively.
The color bars correspond to the \Q{} parameter (top panel), \etak{} parameter (middle panel), and the distance to the fifth nearest neighbor galaxy out to 5 Mpc (bottom panel). 
The total number of galaxies is indicated in each panel.
}
\label{fig:lss_sSFR_Ms}
\end{figure}

To study the role of the large-scale environment,
we perform a bidimensional binned statistic of the datasets
in Figs. \ref{fig:lss_colorMs}--\ref{fig:BPT1sn_4all}. 
We estimate the median of the environmental parameters within each bin. 
We use the same bin number for the two dimensions in these figures. 
We use the routine \verb|binned_statistic_2d| from \textsc{astroML} for this task. 
We show the environmental parameters in color bars (\Q{} in the top panel, \etak{} in the middle panel, and \dfive\ in the bottom panel). 
In this color bar, redder colors mean denser environments. 
The color bar range is approximately centered in the global median value of each environmental parameter.
As expected from the top-left panel of Fig. \ref{fig:lss_Ms}, galaxies tend to be located in denser environments using the \Q{} parameter as the stellar mass increases.
The most massive isolated ellipticals (\mstar\ $> 10^{11}$ \msun) are red galaxies in dense large-scale environments (median \Q{} = -4.6). 
At intermediate masses ($10^{10.5} <$ \mstar\ $< 10^{11}$ \msun), -5.0 and -5.1 are the median values of \Q{} for blue and red galaxies, respectively. 
We note that 16\% of blue isolated ellipticals are in environments denser than \Q{} = -4.6 and, on the other hand, 15\% of red isolated ellipticals of the same mass are in low-density environments with \Q{} $\le$ -6.0.
Therefore, the large-scale environment does not affect the color of the galaxies at these masses directly.
At lower masses (\mstar\ $< 10^{10.5}$ \msun), there is a general trend that bluer galaxies tend to be located in environments of lower densities (i.e., lower \Q{} values). 
However, 38\% of red galaxies of the same mass are also residing in low-density environments of large scale (\Q{} $\le$ -6.0). The median values of \Q{} for blue and red low-mass isolated E galaxies are \mbox{-5.9} and -5.7, respectively.
On the other hand, nearly all the L16 galaxies in the tail of the \Q{} distribution mentioned above are red galaxies.

In the middle panel of Fig. \ref{fig:lss_colorMs}, blue and red galaxies do not 
tend to be located in different environments according to the \etak{} parameter.
Their median values of \etak{} range between -1.2 and -1.4 at different mass bins.
Almost all the isolated E galaxies in the densest environments (\etak{} > -0.4)
are red galaxies. 
Interestingly, a third of them correspond to low-mass galaxies (\mstar\ $< 10^{10.5}$ \msun). 
The fifth nearest neighbor galaxy (bottom panel) of these red low-mass galaxies in dense environments is at a median distance of 1.1 Mpc. As a comparison, red galaxies of higher masses have a median \dfive\ of 2.7 Mpc. Therefore, relatively nearby neighbors may influence the red color of low-mass isolated E galaxies. 
This result could be affected by the tail of L16 galaxies with \dfive{} values lower than 1 Mpc (bottom-right panel of Fig. \ref{fig:lss_Ms}). However, the results are similar if we use only SIG galaxies. Nearly all the SIG galaxies in dense environments are red, 37\% of them with stellar masses lower than $10^{10.5}$ \msun, and the median \dfive\ of the red low-mass galaxies is 1.3 Mpc.
On the other hand, the median \dfive\ of blue low-mass galaxies is similar to that of blue galaxies of higher masses (2.4 and 2.6 Mpc, respectively).

%%%Fig3 

Figure \ref{fig:lss_sSFR_Ms} shows sSFR of isolated E galaxies as a function of their stellar mass. We also use the relation given by \cite{Lacerna+2014} to separate star-forming and quenched galaxies that is
\begin{eqnarray} 
\textrm{log(sSFR)} = -0.65[\textrm{log($M_{\star}$)}-10.31]-10.87  \textrm{ ,} 
\label{eq_sSFR}
\end{eqnarray}
%\\
where sSFR is in units of yr$^{-1}$ and the stellar mass is in units of \msun.
This relation is shown as a black solid line in Fig. \ref{fig:lss_sSFR_Ms}. Most of the isolated E galaxies are quenched galaxies (89\%), that is, they are located below the black line. As well as the case of color$-$stellar mass, we show the large-scale environmental parameters in color bars (\Q{} in the top panel, \etak{} in the middle panel, and \dfive\ in the bottom panel). Again, redder colors mean denser environments in this bar.

The top panel of Fig. \ref{fig:lss_sSFR_Ms} shows that the most massive galaxies are quenched and reside in dense large-scale environments according to the \Q{} parameter. The population of SF isolated ellipticals increases to lower masses. In contrast to the comparison between blue and red galaxies, SF isolated E galaxies are slightly located in lower density environments of large scale 
than quenched isolated ellipticals of the same mass. The median of \Q{} for SF galaxies at masses $10^{10.5} <$ \mstar\ $< 10^{11}$ \msun\ is -5.4, whereas the median of \Q{} is -5.1 for quenched galaxies in the same range of mass, and -6.2 and -5.7, respectively, for the mass range $10^{10} <$ \mstar\ $< 10^{10.5}$ \msun.

For \etak{} (middle panel of Fig. \ref{fig:lss_sSFR_Ms}), the median value for SF isolated galaxies at $10^{10} <$ \mstar\ $< 10^{11}$ \msun\ is -1.5, which is relatively similar to the median value of -1.3 for quenched isolated galaxies of the same mass range. The median distance to the fifth  
nearest neighbor (bottom panel) is 2.8 Mpc for the former, whereas is 2.6 Mpc for the latter. Although in general there are no 
significant differences in the environment of SF and quenched isolated ellipticals using the environmental parameters given by their nearby neighbor galaxies, those galaxies in the densest environments (\etak\ > -0.4) are usually quenched systems (85\% of them). This result is similar 
considering only SIG galaxies.

%%%===================
\subsection{Blue star-forming isolated E galaxies}

A particular sample of isolated E galaxies are those galaxies with blue colors and relatively high sSFR. Both features are not common for very early-type galaxies. There are three blue and SF isolated ellipticals from the L16 sample\footnote{There are four blue, SF isolated E galaxies identified in L16. One of them, UNAM-KIAS 613, is not selected in the current paper after using magnitudes from DR12 data instead of DR7 data as used in L16. We note that, on average, $g$ - $i$ colors from DR7 and DR12 data are very similar.}. They are shown as triangles in Figs. \ref{fig:lss_Ms} -- \ref{fig:lss_sSFR_Ms}. In addition, we have identified other 30 blue and SF isolated E galaxies from the SIG sample (star symbols in Figs. \ref{fig:lss_Ms} -- \ref{fig:lss_sSFR_Ms}). They are both in the blue region of the color--mass diagram and also in the SF region of the sSFR--mass diagram.

The stellar mass range is wider for blue SF ellipticals from the SIG sample than from the L16 sample (10$^{8.9}$ $<$ \mstar/\msun $<$ 10$^{11.1}$ and 10$^{10.0}$ $<$ \mstar/\msun $<$ 10$^{10.3}$, respectively).
At masses $\le$ $10^{10}$ \msun, the blue SF galaxies are 29\% of the isolated elliptical galaxies. This population decrease to 12\% at $10^{10}$ $<$ \mstar/\msun\ $\le$ 10$^{10.5}$ and down to 3\% for masses higher than $10^{10.5}$ \msun. 
They reside in large-scale environments of low density with median \Q{} of -5.9 for SIG and -6.3 for L16 samples. 
For \etak{}, the median values are -1.3 for SIG and -1.1 for L16 samples of blue SF isolated E galaxies.
The median distances to the fifth nearest neighbor galaxy are 2.6 for SIG and 2.3 for L16 samples of blue SF isolated ellipticals. 
Therefore, blue SF isolated E galaxies are more isolated than other isolated ellipticals according to the \Q{} parameter, but they are located on average environments according to the distance to the $k^{th}$ nearest neighbor.
We give more details about this particular subpopulation in the next sections.

%===============================================
\section{BPT diagrams as a function of mass and large-scale environment} 
\label{S4bpt}
%===============================================

%%%FigBPT
\begin{figure*}
\includegraphics[width=9.1cm]{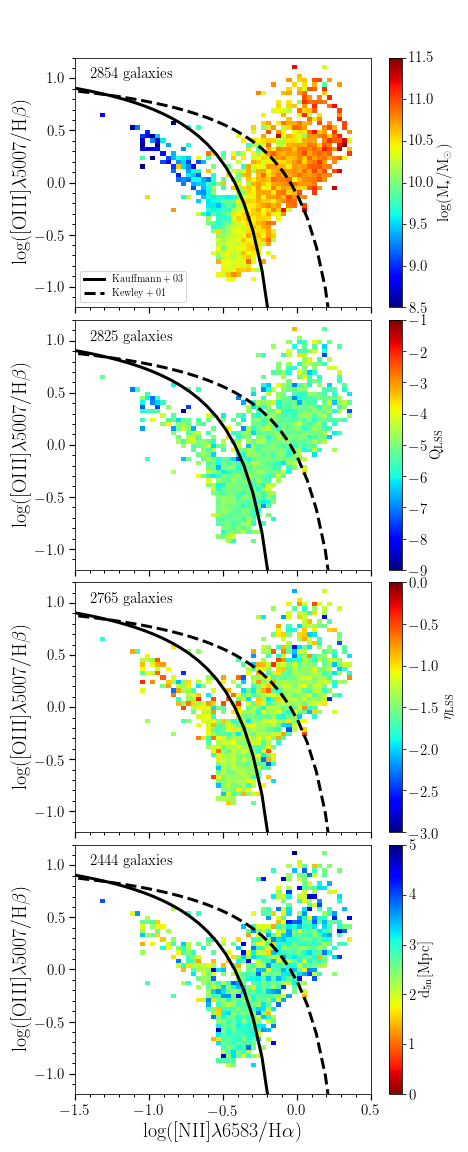}
\includegraphics[width=9.1cm]{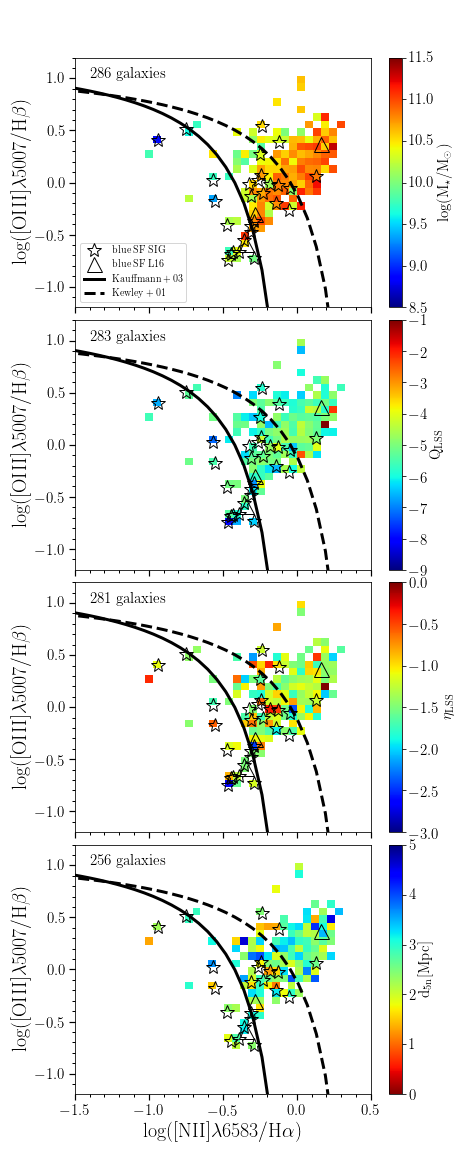}\\
\caption{BPT diagrams for SIG galaxies regardless of the morphology (left column) and for isolated E galaxies (both SIG and L16 samples, right column).
Galaxies are classified as SFN
to the left of the solid line of \citet[][]{Kauffmann+2003},
transition objects between the solid line and the dashed line of \citet[][]{Kewley+2001},
or AGN to the right of the dashed line.
The black stars and black triangles in the right column show the blue SF isolated Es from the SIG and L16 samples, respectively.
The color bars, from top to bottom, correspond to the stellar mass, \Q{}, \etak{}, and the distance to the fifth nearest neighbor galaxy out to 5 Mpc. 
The total number of galaxies is indicated in each panel.
} 
\label{fig:BPT1sn_4all}
\end{figure*} 

The BPT diagram, initially introduced by \citet{BPT1981} and redefined by \citet{VeilleuxOsterbrock1987}, is an optical diagnosis for nuclear classification of production mechanisms for ionization and emission lines. 
The mechanisms are usually produced either by photoionization by massive OB stars (regions with star formation) or nonstellar sources such as active galactic nuclei (AGN). 
For a 
more in-depth analysis
of the AGN and star-forming nuclei (SFN) 
percentages in isolated galaxies, 
see a companion paper \citep{Argudo-Fernandez+2018}.

The top-left panel of Fig. \ref{fig:BPT1sn_4all} shows the BPT diagram 
for SIG isolated galaxies regardless of the morphology as a function
of stellar mass. We find that isolated galaxies follow the known trend that distinguishes low-mass galaxies and high-mass galaxies in this diagram, that is, more massive galaxies tend to be located diagonally toward the right, 
whereas lower mass galaxies tend to be located diagonally toward the left. 
Here we define as AGN and SFN those galaxies that fall above the relation of \cite[][dashed line]{Kewley+2001} and below the relation of \citet[][solid line]{Kauffmann+2003}, respectively. Thus, the most massive galaxies are mostly located in the AGN region, and the low-mass galaxies tend to be located in the SFN region.
We define the galaxies falling between both relations as `transition objects' (TO)\footnote{The transition objects are also known as `SFN+AGN composite' galaxies. However, their emission lines may not be dominated by SFN nor an AGN (e.g., see discussion in \citealt{CidFernandes+2010} and \citealt{McIntosh+2014}).}.

The other left panels in the figure, from top to bottom, show the BPT diagram for SIG isolated galaxies regardless of the morphology as a function of the \Q, \etak, and \dfive\ parameters. In our understanding, it is the first time that the BPT diagram is used as a function of large-scale environment for ``field'' galaxies. In contrast to the stellar mass, there is not a 
robust dependence of the BPT diagram on the environmental parameters used in this work. There is a mild trend that the galaxies located diagonally toward the left in the SFN region are in slightly denser environments than the other galaxies using \etak\ and \dfive. 
The median \dfive\ is 2.3 Mpc for the branch toward the left, whereas it is 2.6 Mpc for galaxies in the SFN region which are part of the other branch toward the right.
On the other hand, the median Q is -5.2 for the former, which is slightly lower (i.e., less dense) than for the latter (Q = -5.0). 
The isolated galaxies in the branch toward the left are probably close to structures formed by many low-mass galaxies, which are mostly blue, and 
with a considerable amount of gas available to form stars within these structures.

The right column shows the same plots as the left column, but 
for isolated E galaxies. They are 286 galaxies that satisfy the conditions: i) positive emission lines 
of \ion{O}{iii}, \ion{N}{ii}, H${\beta}$ and H${\alpha}$,
and ii) SNR of these emission lines $\ge$ 3.  
Most of them are located diagonally toward the right on the diagram. The highest 
percentage corresponds to AGN (64\%), 
followed by transition objects (28\%). 
The minority is SFN (8\%).
There is a trend toward lower \Q\ values, that is, lower density environments, from AGN to SFN regions. The median \Q\ is -5.0 for AGN, -5.3 for transition objects, and -5.6 for SFN. We do not find these trends using \etak\ (-1.3 as the median value for all the regions in the BPT diagram) and \dfive\ (medians between -2.5 and -2.7 Mpc for all the regions). The results point out that the isolated E galaxies in the SFN region might be located in a large group of low-mass galaxies with vast distances between them.

There are 31 (94\%) blue SF E galaxies that satisfy the conditions to be shown in the BPT diagram. They mostly follow the trend of isolated E galaxies, but with a particular concentration in both the SFN and TO regions (45\% and 42\%, respectively). 
Therefore,
at least a half of the blue SF isolated E galaxies selected with both integrated optical $g-i$ color and sSFR from the nuclear region are 
undoubtedly star-forming galaxies. 
They reside
in low-density environments using \Q\ with \mbox{-5.7} as the median value but 
in standard environments 
using \etak\ with \mbox{-1.2} as the median value. 
These galaxies can be 
exceptional cases of being located in the vicinity of groups of low-mass galaxies separated at large distances. Probably, the gas within and around these structures is not as hot as the gas in galaxy clusters, which facilitates star formation.

%========================
\section{Discussion} 
\label{Sdiscussion}
%========================

The results 
here and in L16 point out that a fraction of E galaxies can be different from the red and quenched classical E galaxies 
regarding optical colors and sSFR. At fixed stellar mass, this fraction decrease with the local environment from isolation to cluster of galaxies. This fraction partially decrease with denser environments defined within a radius of 5 Mpc. 

L16 find that the four blue SF isolated ellipticals, from the UNAM-KIAS sample, exhibit radial color gradients with bluer colors toward the galaxy center, whereas 
most of the red or quenched ellipticals exhibit negative or flat radial color profiles. 
The positive color gradient may be evidence of dissipative infall of cold gas, which promotes recent star formation in the central regions. The latter may imply a rejuvenation process produced by recent 
star formation in some early-type galaxies 
\citep{Treu+2005, Thomas+2010, Haines+2015}.

We have found 33 (5.8\%) blue SF E galaxies in the SIG and L16 samples. 
They are found in very low-density environments using the \Q{} parameter, where $\sim90\%$ of the blue SF ellipticals have values of \Q\ $\le$ -5.0. This result supports the picture that isolated environments seem to propitiate the rejuvenation of E galaxies. 
However, 80\% of the galaxies with \Q\ $\le$ -5.0 correspond to classical `red and dead' E galaxies. 
Therefore, the large-scale environment is not playing the primary role to determine the integrated properties of isolated E galaxies such as color and sSFR. 

Furthermore, when we use the parameters \etak\ y \dfive, the blue SF isolated E galaxies do not appear to be located 
mainly in low-density environments. 
Around 50\% of them have values of \etak\ $\le$ -1.3, which is the median value of this parameter for all the isolated E galaxies, and 55\% of the blue SF galaxies have the fifth nearest neighbor galaxy at distances of \dfive\ $\ge$ 2.5 Mpc.

If the large-scale environment plays some role in the integrated properties of a fraction of E galaxies, this is true until some mass scale where processes related directly with either stellar mass or the mass of the host dark matter halo are the 
dominant factors that regulate the star formation in galaxies.  
In other words, over this mass scale, the local galaxy processes within the virial radius rather than external processes related with the large-scale environments are mostly responsible for the quenching and general properties of E galaxies. 
What is the value of this mass scale? 
The large-scale environment seems to have a negligible role from a stellar mass scale around $10^{10.6}$ \msun.
Only two percent of the galaxies with \ms > $10^{10.6}$ \msun\ are blue SF isolated ellipticals, whereas 
they are twelve percent 
below this mass.
\citet{Argudo-Fernandez+2018} use the same value of stellar mass to separate low-mass and high-mass isolated galaxies,
in which the
fraction of AGN for early-type, red, and quenched SIG galaxies matches the fraction of AGN for late-type, blue, and star-forming SIG galaxies. They find that once the galaxy reaches a stellar mass about \mstar\ $\sim$ $10^{10.6}$ \msun, the probability that an isolated galaxy hosts an AGN is independent of its morphology, color, or sSFR. 
Therefore, it is very likely that the minimum role of the large-scale environment at higher masses is because of the dominant presence of AGN in these galaxies
\citep[see also][]{Hirschmann+2013}.

We can see a partial influence of the large-scale environment on galaxies with \mstar\ $\lesssim$ $10^{10.6}$ \msun\ because 3\% of them are blue SF isolated ellipticals located in environments with \Q\ $>$ -5.0, whereas 14\% are blue SF galaxies with \Q $<$ -5.0. However, as mentioned above, most of the galaxies below this limit in \Q\ are red and quenched ellipticals, including the low-mass isolated galaxies. Therefore, the processes of cooling and infall of gas from large scales are very inefficient in ellipticals.

Another scenario is that some E galaxies are recycling the remaining gas within their virial radius to form new stars, which is more plausible in low-mass galaxies in low-density environments. 
This process is also inefficient because of the low fraction of 
low-mass blue SF galaxies.

Interestingly, the 
percentage of AGN is relatively constant with mass for isolated early-type galaxies \citep{Argudo-Fernandez+2018}.
They find that about $\sim$50\% of isolated galaxies with early-type morphologies are AGN or transition objects, at least for stellar masses down to $10^{10}$ \msun.
Perhaps, in addition to other feedback mechanisms, AGN is also an 
essential factor to produce the cessation of star formation and gas accretion so efficiently in isolated low-mass E galaxies.
Evidence of AGN feedback has been found even for galaxies with stellar masses as low as $\sim 10^{9.5}$ \msun\ \citep{Penny+2018}.    
The BPT diagrams show that isolated E galaxies are mostly AGNs. If we restrict to galaxies with stellar masses $\lesssim$ $10^{10.6}$ \msun,  
we find 56\% 
of them are AGN, 29\% 
are transition objects, and 15\% 
are SFN. 
Therefore, AGN is also dominant for low-mass isolated E galaxies.
Furthermore, 65\% of the red low-mass E galaxies and 74\% of the quenched low-mass E galaxies are AGN in the BPT diagram. 
In the densest large-scale environments, the low-mass isolated E galaxies are red and quenched, which is an indication that dense environments at both local and large scales (e.g., galaxy clusters and dense filaments, respectively) affect the color and SF of galaxies. However, only a small fraction of the red and dead low-mass E galaxies are located in the vicinity of these dense structures (for instance, 8\% of the red or quenched low-mass galaxies have \Q\ $\ge$ -4.6, and 10\% of the red or quenched low-mass galaxies have \etak\ > -0.6). Therefore, AGN is a more 
critical factor in the red color and quenched activity of low-mass isolated E galaxies than the large-scale environment.
The results do not change when we also include WLGs (see Appendix \ref{Ap_wlg}).

\citet{Rosito+2018} studied field spheroid-dominated systems from a hydrodynamical cosmological simulation \citep{PedrosaTissera2015} that included star formation, chemical evolution, and Supernova feedback. They find that the simulated galaxies with masses 10$^{9.4}$ < \mstar\ < 10$^{10.8}$ \msun\ are in general bluer and with higher star formation rates than observed isolated early-type galaxies. Therefore, some other efficient mechanisms for avoiding 
another disc growth or
for quenching the star formation seem to be necessary. AGN could be one of these mechanisms, not only for massive systems as is usually invoked but also for low-mass early-type galaxies.

%========================
\section{Summary and conclusions} 
\label{Sconclusions}
%========================

We have quantified the role of the large-scale environment as a function of stellar mass in integrated physical properties such as color and sSFR of nearby isolated galaxies with E morphologies. For this, we separated galaxies between red or blue and quenched or star-forming systems. We use three environmental estimators of the large-scale structure within a projected radius of 5 Mpc around isolated E galaxies.
They are the tidal strength parameter \Q,  the projected density to the $k^{th}$ nearest neighbor \etak, and the distance to the fifth nearest neighbor galaxy \dfive.

The most massive galaxies (\mstar\ $>$ 10$^{11}$ \msun) are red or quenched galaxies that reside in dense large-scale environments estimated via \Q. 
At lower masses, we find that \Q\ does not affect the color of galaxies directly
because 16\% of blue isolated ellipticals are in very dense environments (\Q $\ge$ -4.6), 
whereas 15\% of red low-mass galaxies are residing in very low-density environments (\Q $\le$ -6.0) at 10$^{10.5}$ < \ms\ < 10$^{11}$ \msun. 
The median values of \Q\ for blue and red low-mass (\mstar\ $<$ 10$^{10.6}$ \msun) isolated E galaxies are -5.9 and \mbox{-5.6}, respectively. 
Blue and red galaxies do not tend to be located in different environments according to the \etak\ parameter. 
Almost all the isolated E galaxies in the densest large-scale environments with this parameter (e.g., \etak\ > -0.4) are red galaxies, where a third of them are low-mass galaxies. Our results suggest that nearby neighbors (with median \dfive\ $\le$ 1.3 Mpc) may influence the red color of low-mass isolated E galaxies. 

The population of SF isolated ellipticals increases to lower masses. SF isolated E galaxies are slightly located in lower density environments of large scale than quenched isolated ellipticals of the same mass. Median \Q\ is -6.2 for the former, whereas it is -5.7 for the latter in the mass range $10^{10}$ $<$ \ms $<$ $10^{10.5}$ \msun. 
On the other hand, those galaxies in the densest environments are usually quenched systems (85\% of them). 

We have identified 33 blue, SF isolated ellipticals. They correspond to $\sim$6\% of the sample with sSFR data available. They are located in lower density environments than other isolated ellipticals according to the \Q{} parameter (median \Q $\sim$ -6.0 compared with $\sim$ -5.1 for the full sample), but they are located on average large-scale environments according to the distance to the $k^{th}$ nearest neighbors.
On the other hand, 80\% of the galaxies with \Q\ $\le$ -5.0 correspond to classical `red and dead' E galaxies. 
The overall results suggest
that the large-scale environment is not playing the primary role to determine the integrated properties of isolated E galaxies such as color and sSFR.

We also showed the BPT diagram as a function of the large-scale environment. In contrast to the stellar mass, there is not a 
robust dependence of the BPT diagram on the environmental parameters used in this work. There is a mild trend that the isolated galaxies, regardless of their morphology, located in the left branch of the SFN region are 
close to structures with a considerable amount
of gas available to form stars.
For isolated galaxies with E morphologies,
the highest percentage of SLGs
corresponds to AGN (64\%), followed by transition objects (28\%). On the other hand, we find that half of the blue SF isolated E galaxies are SFN. 

Our results point out that the large-scale environment seems to be negligible from a mass scale around $10^{10.6}$ \msun.
It is very likely that the minimum role of the large-scale environment at higher masses is because of the dominant presence of AGN. For lower masses, the processes of cooling and infall of gas from large scales are very inefficient in ellipticals (e.g., the low fraction of blue SF low-mass galaxies). We discussed that optical
AGN is also an essential factor to keep most of the low-mass E galaxies red or quenched.

\begin{acknowledgements}
We thank the anonymous referee for the constructive revision of our paper.
IL acknowledges partial financial support from PROYECTO FONDECYT REGULAR 1150345.
MAF is grateful for financial support from the CONICYT Astronomy Program CAS-CONICYT project No.\,CAS17002 and from CONICYT FONDECYT project No.\,3160304. SDP acknowledges financial support from the Spanish Ministerio de Economía y Competitividad under grant AYA2013-47742-C4-1-P, AYA2017-79724-C4-4-P from the Spanish PNAYA, and from Junta de Andalucía Excellence Project PEX2011-FQM-7058.

This research made use of \textsc{astropy}, a community-developed core \textsc{python} ({\tt http://www.python.org}) package for Astronomy \citep{2013A&A...558A..33A}; \textsc{ipython} \citep{PER-GRA:2007}; \textsc{matplotlib} \citep{Hunter:2007}; \textsc{numpy} \citep{:/content/aip/journal/cise/13/2/10.1109/MCSE.2011.37}; and \textsc{scipy} \citep{citescipy}.

Funding for SDSS-III has been provided by the Alfred P. Sloan Foundation, the Participating Institutions, the National Science Foundation, and the U.S. Department of Energy Office of Science. The SDSS-III web site is http://www.sdss3.org/. 
SDSS-III is managed by the Astrophysical Research Consortium for the Participating Institutions of the SDSS-III Collaboration including the University of Arizona, the Brazilian Participation Group, Brookhaven National Laboratory, University of Cambridge, University of Florida, the French Participation Group, the German Participation Group, the Instituto de Astrofisica de Canarias, the Michigan State/Notre Dame/JINA Participation Group, Johns Hopkins University, Lawrence Berkeley National Laboratory, Max Planck Institute for Astrophysics, New Mexico State University, New York University, Ohio State University, Pennsylvania State University, University of Portsmouth, Princeton University, the Spanish Participation Group, University of Tokyo, University of Utah, Vanderbilt University, University of Virginia, University of Washington, and Yale University.

\end{acknowledgements}

\bibliographystyle{aa}
\bibliography{references}
%\end{document} 

\begin{appendix}

\section{BPT including weak emission-line galaxies}
\label{Ap_wlg}

In the manuscript, we use galaxies with an SNR $\ge$ 3 for each line in the BPT diagram. We refer to them as SLGs. Here, we explore the case of also including WLGs in the results presented in Sect. \ref{S4bpt}.
According to \citet{CidFernandes+2010}, these galaxies must be detected with both H$\alpha$ and [NII]$\lambda$6584 lines with SNR $\ge$ 3, but there are different criteria of SNR for H$\beta$ and  [OIII]$\lambda$5007 emission lines. We make the following definitions 
to consider both SLGs and WLGs.

(i) SLG+WLO, which includes weak [OIII]$\lambda$5007 line (i.e., SNR([OIII]$\lambda$5007) $\geq$ 0, SNR(H$\beta$) $\geq$3, SNR(H$\alpha$) $\geq$ 3, and SNR([NII]$\lambda$6584) $\geq$ 3).

(ii) SLG+WLH, which includes weak H$\beta$ line (i.e., SNR(H$\beta$) $\geq$ 0, SNR([OIII]$\lambda$5007) $\geq$ 3, SNR(H$\alpha$) $\geq$ 3, and SNR([NII]$\lambda$6584) $\geq$ 3).

(iii) SLG+WLOH, which includes both weak [OIII]$\lambda$5007 and H$\beta$ lines (i.e., SNR([OIII]$\lambda$5007) $\geq$ 0, SNR(H$\beta$) $\geq$ 0, SNR(H$\alpha$) $\geq$ 3, and SNR([NII]$\lambda$6584) $\geq$ 3). 

Also, we define as no reliable line galaxies (NRGs) those isolated elliptical galaxies where the SNR for H$\alpha$, H$\beta$, [OIII]$\lambda$5007, and [NII]$\lambda$6584 emission lines is greater or equal than 0. In all these cases, the flux of each emission line must be positive.

For only SLG, the percentages reported in Sect. \ref{S4bpt} are 64\% as AGN, 28\% as TO, and 8\% as SFN. These values are similar to the percentages for the cases SLG+WLO, SLG+WLH, and SLG+WLOH: 63--69\% as AGN, 24--29\% as TO, and 7--8\% as SFN.
For the case of NRG, the percentage of AGN decreases to 61\%, for TO is 24\%, and for SF increases to 15\%.
We detail the results in Table \ref{tabla1}, where AGN is separated in Seyfert and low-ionization nuclear emission-line region (LINER) galaxies using the criterion described in \citet{Schawinski+2007}.
AGN is dominated by LINERs for all the types of emission lines galaxies, although this is less strong 
for NRG.

%%% TABLA1
\begin{table*}
\caption{Percentages and number of galaxies according to the locations in the BPT diagram for different SNR criteria of emission lines.
}
\centering
\begin{tabular}{c c c c c c c c c c c}
\hline
\hline
type  &	 \%SFN	&  \%TO	 &  \%Sy & \%LINER &		SFN	& TO  & Sy  &  LINER &  Total \\
\hline
SLG	    & 8.4	& 28.0	& 8.7 &	54.9 	& 24 &	80 &	25 &	157 &	286	\\
SLG+WLO & 8.2 &	28.8	& 8.6 &	54.5	 &	 24 &	84 &	25 &	159	& 292	\\
SLG+WLH	& 7.1 &	24.3 &	16.3 &	52.4		& 24	& 82	& 55	& 177	& 338	\\
SLG+WLOH &	 6.9 &	24.5 &	15.8 &	52.8  &	 25 &	88 &	57 &	190 &	360	\\
NRG &	14.9	& 23.9 &	17.3 &	43.9  &	 74 &    119 &	86 &	218 &	497\\
\hline
\end{tabular}

\tablefoot{Columns are the type of emission-line galaxies and the percentage of SFN, TO, Seyfert, and LINER galaxies. Also, the number of SFN, TO, Seyfert, and LINER galaxies followed by the total number of isolated E galaxies of each type.
}
\label{tabla1}
\end{table*}

We also reported that the median \Q\ is -5.0 for AGN of only SLG. This median value is the same for AGN of SLG+WLO, SLG+WLH, SLG+WLOH, and NRG samples. Median \Q\ for transition objects and SFN of only SLG are very similar to the median values of the other samples (around -5.3 and -5.6, respectively). 
There are differences
if AGN is separated in Seyfert and LINER. The median \Q\ of LINERs is -5.0 for all the samples, but decreases to -5.7 for Seyferts of only SLG and SLG+WLO. The median \Q\ of Seyferts for SLG+WLH, SLG+WLOH, and NRG is about -5.2. We detail these results in Table \ref{tabla2}.
Therefore, there is a general trend that high-ionized galaxies (i.e., Seyferts) are located in regions of lower densities as well as SF isolated E galaxies as measured with the \Q\ estimator. 
On the other hand, LINERs have been proposed as `retired' galaxies, that is,  strongly dominated by hot old stars and not related with nuclear ionization \citep{Stasinska+2008,Singh+2013}. Here we do not discuss the nature of LINERs, but we note they dominate the population of isolated elliptical galaxies and reside in average environments of large scale.

%%% TABLA2
\begin{table*}
\caption{Median \Q\ and number of galaxies according to the locations in the BPT diagram for different SNR criteria of emission lines.
}
\centering
\begin{tabular}{c c c c c c c c c c c}
\hline
\hline
type  &	 \Q\ SFN	&  \Q\ TO	 &  \Q\ Sy & \Q\ LINER &		SFN	& TO  & Sy  &  LINER &  Total \\
\hline
SLG &	-5.6 &	-5.3 &	-5.7 &	-5.0 &	24 &	78 &	25	&  156 &	  283	\\
SLG+WLO	& -5.6	& -5.2 &	-5.7 &	-5.0 &	24 &	82 &	25 &	  158 &	  289	\\
SLG+WLH	& -5.6 &	-5.2 &	-5.2 &	-5.0 &	24 &	80 &	55 &	  176 &	  335	\\
SLG+WLOH &	-5.5 &	-5.1 &	-5.2 &	-5.0 & 	25 &	86 &	57 &	  189 &	  357 \\	
NRG	& -5.3 &	-5.2 &	-5.3 &	-5.0 &	74 &    117 &	86 &	  217	&   494 \\
\hline
\end{tabular}

\tablefoot{Columns are the type of emission-line galaxies and the median value of \Q\ for SFN, TO, Seyfert, and LINER galaxies. Also, the number of SFN, TO, Seyfert, and LINER galaxies followed by the number of isolated E galaxies of each type with estimation of \Q.
}
\label{tabla2}
\end{table*}

For \etak, it is still true that the median is about -1.3 for all the samples of SFN and transition objects. If AGN is separated in Seyfert and LINER, the median is between -1.2 and -1.3 for Seyferts and between -1.3 and -1.4 for LINERs of only SLG, SLG+WLO, SLG+WLH, SLG+WLOH, and NRG samples. 
We show these results in Table \ref{tabla3}.

%%% TABLA3
\begin{table*}
\caption{Median \etak\ and number of galaxies according to the locations in the BPT diagram for different SNR criteria of emission lines.
}
\centering
\begin{tabular}{c c c c c c c c c c c}
\hline
\hline
type  &	 \etak\ SFN	&  \etak\ TO	 &  \etak\ Sy & \etak\ LINER &		SFN	& TO  & Sy  &  LINER &  Total \\
\hline
SLG	& -1.3 &	-1.3 &	-1.2 &	-1.3 &	24	 & 78 &	24	& 155 &	281	\\
SLG+WLO	& -1.3 &	-1.3 &	-1.2 &	-1.3 &	24 &	82 &	24 &	157 &	287	\\
SLG+WLH	& -1.3 &	-1.3 &	-1.2 &	-1.4 &	24 &	80 &	54 &	174 &	332	\\
SLG+WLOH &	-1.3 &	-1.3 &	-1.2 &	-1.4 &	25 &	86 &	56 &	187 &	354	\\
NRG &	-1.3 &	-1.2 &	-1.4 &	-1.4 &	74 &     117 &	85 &	215 &	491 \\
\hline
\end{tabular}

\tablefoot{Columns are the type of emission-line galaxies and the median value of \etak\ for SFN, TO, Seyfert, and LINER galaxies. Also, the number of SFN, TO, Seyfert, and LINER galaxies followed by the number of isolated E galaxies of each type with estimation of \etak.
}
\label{tabla3}
\end{table*}

For \dfive, we obtain medians between -2.4 and \mbox{-2.7} Mpc for all the regions in all the samples with a slight trend of smaller values for Seyferts and higher values for SFNs. 
We detail these results in Table \ref{tabla4}.

%%% TABLA4
\begin{table*}
\caption{Median \dfive\ and number of galaxies according to the locations in the BPT diagram for different SNR criteria of emission lines.
}
\centering
\begin{tabular}{c c c c c c c c c c c}
\hline
\hline
type  &	 \dfive\ SFN	&  \dfive\ TO	 &  \dfive\ Sy & \dfive\ LINER &		SFN	& TO  & Sy  &  LINER &  Total \\
\hline
SLG	& 2.7 &		2.5 &		2.4 &		2.6 &		21 &	67 &	22 &	146 &	256	\\
SLG+WLO &	2.7	&	2.5	&	2.4 &		2.6 &		21 &	71 &	22 &	148 &	262	\\
SLG+WLH &	2.7	&	2.5 &		2.5 &		2.7	&	21 &	69 &	49 &	161 &	300	\\
SLG+WLOH &	2.7 &		2.4 &		2.5 &		2.7 &		22 &	74 &	51 &	171 &	318	\\
NRG &	2.5 &		2.4 &		2.6 &		2.7 &		67 &	105 &	77 &	196 &	445 \\
\hline
\end{tabular}

\tablefoot{Columns are the type of emission-line galaxies and the median value of \dfive\ for SFN, TO, Seyfert, and LINER galaxies. Also, the number of SFN, TO, Seyfert, and LINER galaxies followed by the number of isolated E galaxies of each type with estimation of \dfive.
}
\label{tabla4}
\end{table*}

In the Discussion of Sect. \ref{Sdiscussion}, it is reported that 56\% of the low-mass isolated elliptical galaxies are classified as AGN for only SLG. 
It is the same percentage of AGN for SLG+WLO and slightly higher for SLG+WLH and SLG+WLOH (60\% and 61\%, respectively).
If AGN is separated in Seyfert and LINER, the percentage of the former is typically around 20\%, whereas the percentage of the latter is around 38\%. Table \ref{tabla5} shows the results in detail.

%%% TABLA5
\begin{table*}
\caption{Percentages and number of low-mass isolated E galaxies (\mstar\ $\le$ $10^{10.6}$ \msun) according to the locations in the BPT diagram for different SNR criteria of emission lines.
}
\centering
\begin{tabular}{c c c c c c c c c c c}
\hline
\hline
type  &	 \%SFN	&  \%TO	 &  \%Sy & \%LINER &		SFN	& TO  & Sy  &  LINER &  Total \\
\hline
SLG &	15.0 &	29.2 &	17.5 &	 38.3 &	 18 &	35 &	21 &	46 &	120	\\
SLG+WLO &	14.9 &	28.9 &	17.4 &	 38.8 &		 18 &	35 &	21 &	47 &	121	\\
SLG+WLH	& 13.6 &	26.5 &	22.7 &	 37.1 & 18 &	35 &	30 &	49 &	132	\\
SLG+WLOH &	13.2 &	25.7 &	23.5 &	 37.5 &	 18 &	35 &	32 &	51 &	136	\\
NRG &	19.5 &	26.8 &	24.7 &	 28.9 &	 	 37 &	51 & 	47 &	55 &	190\\
\hline
\end{tabular}

\tablefoot{Columns are the type of emission-line galaxies and the percentage of SFN, TO, Seyfert, and LINER galaxies. Also, the number of SFN, TO, Seyfert, and LINER galaxies followed by the total number of low-mass galaxies of each type.
}
\label{tabla5}
\end{table*}

Furthermore, we reported that 65\% of the red low-mass E galaxies are AGN in the only SLG sample. This percentage is the same for SLG+WLO and increases to 69\% for both SLG+WLH and SLG+WLOH. It decreases to 58\% for NRG.
If AGN is separated in Seyfert and LINER, the percentage of the former ranges between 20\% and 27\%, whereas the percentage of the latter is around 43\%, except for NRG where the percentage is 31\%. We also reported that 74\% of the quenched low-mass E galaxies are AGN for only SLG. This percentage is also consistent for SLG+WLO and increases to 77\% for SLG+WLH and 78\% for SLG+WLOH. It decreases to 63\% for NRG.
If AGN is separated in Seyfert and LINER, the percentage of the former ranges between 22\% and 30\%, whereas the percentage of the latter ranges between 48\% and 52\%, except for NRG where the percentage is 34\%.

%%% TABLA6
\begin{table*}
\caption{Percentages and number of red, quenched, blue, and SF low-mass isolated E galaxies (\mstar\ $\le$ $10^{10.6}$ \msun) according to the locations in the BPT diagram for different SNR criteria of emission lines.
}
\centering
\begin{tabular}{c c c c c c c c c c c}
\hline
\hline
type  &	 \%SFN	&  \%TO	 &  \%Sy & \%LINER &		SFN	& TO  & Sy  &  LINER &  Total \\
\hline
red	SLG &	 6.5 &	29.0 &	20.4 &	44.1	&	  6 &	27 &	19 &	41 &	93	\\
red	 SLG+WLO &	 6.5	& 29.0 &	20.4 &	44.1 &	  6 &	27 &	19 &	41 &	93	\\
red	 SLG+WLH &	 5.7 &	25.7 &	26.7 &	41.9 &	  6 &	27	& 28	 & 44 &	105	\\
red	SLG+WLOH &	 5.6 &	25.2 &	27.1 &	42.1 &	 6 &	27 &	29 &	45 &	107	\\
red	 NRG &	15.6 &	26.9 &	26.9 &	30.6 &	 25 &	43 &	43 &	49 &	160 \\
quenched SLG &	 1.2 &	24.7 &	22.4 &	51.8 &	 	  1 &	21 &	19 &	44 &	85	\\
quenched	 SLG+WLO &	 1.2 &	24.4 &	22.1 &	52.3 &		  1 &	21 &	19 &	45 &	86	\\
quenched	 SLG+WLH &	 1.0 &	21.6 &	28.9 &	48.5 &	   1 &	21 &	28 &	47 &	97	\\
quenched	SLG+WLOH &	 1.0 &	20.8 &	29.7 &	48.5 &  1 &	21 &	30 &	49 &	101	\\
quenched	 NRG	& 12.9 &	23.9 &	29.0 &	34.2 &	   20 &	37 &	45 &	53 &	155 \\
blue	 SLG	& 44.4 &	29.6 &	 7.4 &	18.5 &		 12 &	 8 &	 2 &	 5 &	27	\\
blue	 SLG+WLO &	42.9 &	28.6	& 7.1 &	21.4 &	 12 &	 8 &	 2 &	 6 &	28	\\
blue	 SLG+WLH &	44.4 &	29.6 &	 7.4	& 18.5 &	 12 &	 8 &	 2 &	 5 &	27	\\
blue	SLG+WLOH &	41.4 &	27.6 &	10.3 &	20.7 &	 	 12 &	 8 &	 3 &	 6 &	29	\\
blue	 NRG &	40.0 &	26.7 &	13.3 &	20.0 &	 12 &	 8 &	 4 &	 6 &	30 \\
SF	 SLG &	48.6 &	40.0 &	 5.7 &	 5.7 &		 17 &	14 &	 2 &	 2 &	35	\\
SF	 SLG+WLO &	48.6 &	40.0 &	 5.7 &	 5.7 &  17 &	14 &	 2 &	 2	& 35	\\
SF	 SLG+WLH &	48.6 &	40.0 &	 5.7 &	 5.7 &	 17 &	14 &	 2 &	 2 &	35	\\
SF	SLG+WLOH &	48.6 &	40.0 &	 5.7 &	 5.7 & 17 &	14 &	 2 &	 2 &	35	\\
SF	 SLG+NRG &	48.6 &	40.0 &	 5.7 &	 5.7 &	 17 &	14 &	 2 &	 2	& 35 \\
\hline
\end{tabular}

\tablefoot{Columns are the type of emission-line galaxies and the percentage of SFN, TO, Seyfert, and LINER galaxies. Also, the number of SFN, TO, Seyfert, and LINER galaxies followed by the total number of low-mass galaxies of each type.
}
\label{tabla6}
\end{table*}

Therefore, our claim that AGN is a more critical factor in the red color and quenched activity of low-mass isolated E galaxies than the large-scale environment is still valid when we also include weak emission lines.

\section{LSSGalPy}
\label{Ap_lss} 

LSSGalPy\footnote{Available at \texttt{https://github.com/margudo/LSSGALPY}} \citep{Argudo-Fernandez+2015b,Argudo-Fernandez+2017} is a tool for the interactive visualization of the 3D positions (right ascension, declination, and redshift) of  
galaxy samples 
for the locations 
in their local and large-scale environments. Using this tool, \citet{Argudo-Fernandez+2015} find that most of the SIG galaxies are mostly distributed following the LSS (related more to the outer parts of filaments, walls, and clusters), and only one-third of SIG galaxies are located in voids. Moreover, \citet{Argudo-Fernandez+2016} explored the relation of the SIG with respect the LSS as a function of the $\rm Q_{LSS}$ parameter. They find that isolated galaxies with lower values of the $\rm  Q_{LSS}$ are mainly located in voids or low-density regions, and isolated galaxies with higher $\rm Q_{LSS}$ are more closely related to denser structures, such as the filaments or walls defining the LSS of the Universe.

\begin{sidewaysfigure*}
\includegraphics[width=24.6cm]
{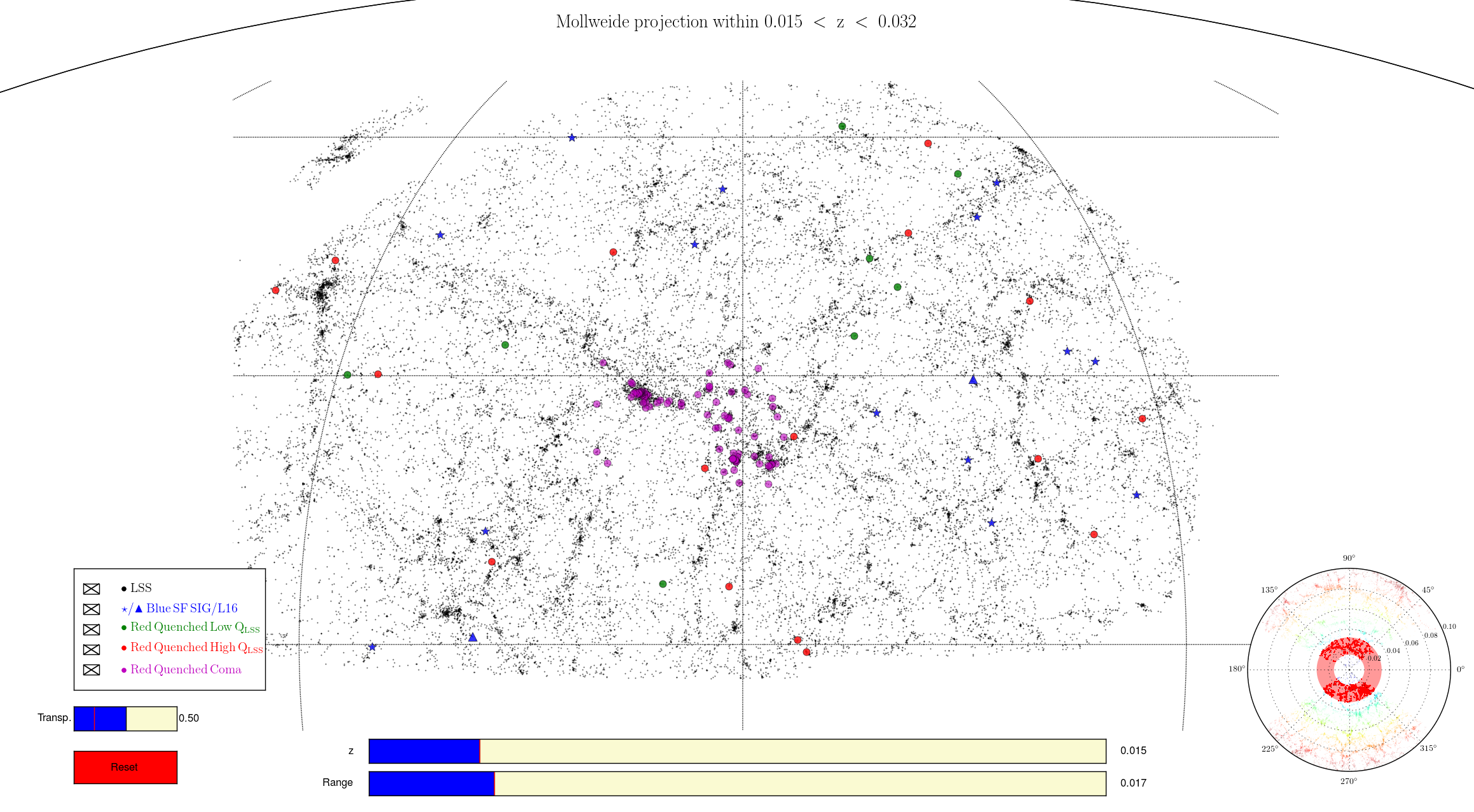}
\caption{Mollweide projection of SDSS galaxies at $0.015 < z < 0.032$. Blue SF isolated E galaxies from SIG and L16 samples are shown as blue stars and blue triangles, respectively. 
Red, quenched isolated E galaxies in environments of high (\Q\ $\ge$ -4.75) and low (\Q\ $\le$ -7.0) density are shown in red and green circles, respectively.
As a comparison, E galaxies of the Coma supercluster from L16 are shown in magenta circles. 
}
\label{fig:mollweide}
\end{sidewaysfigure*}

\begin{sidewaysfigure*}
\includegraphics[width=24.6cm]
{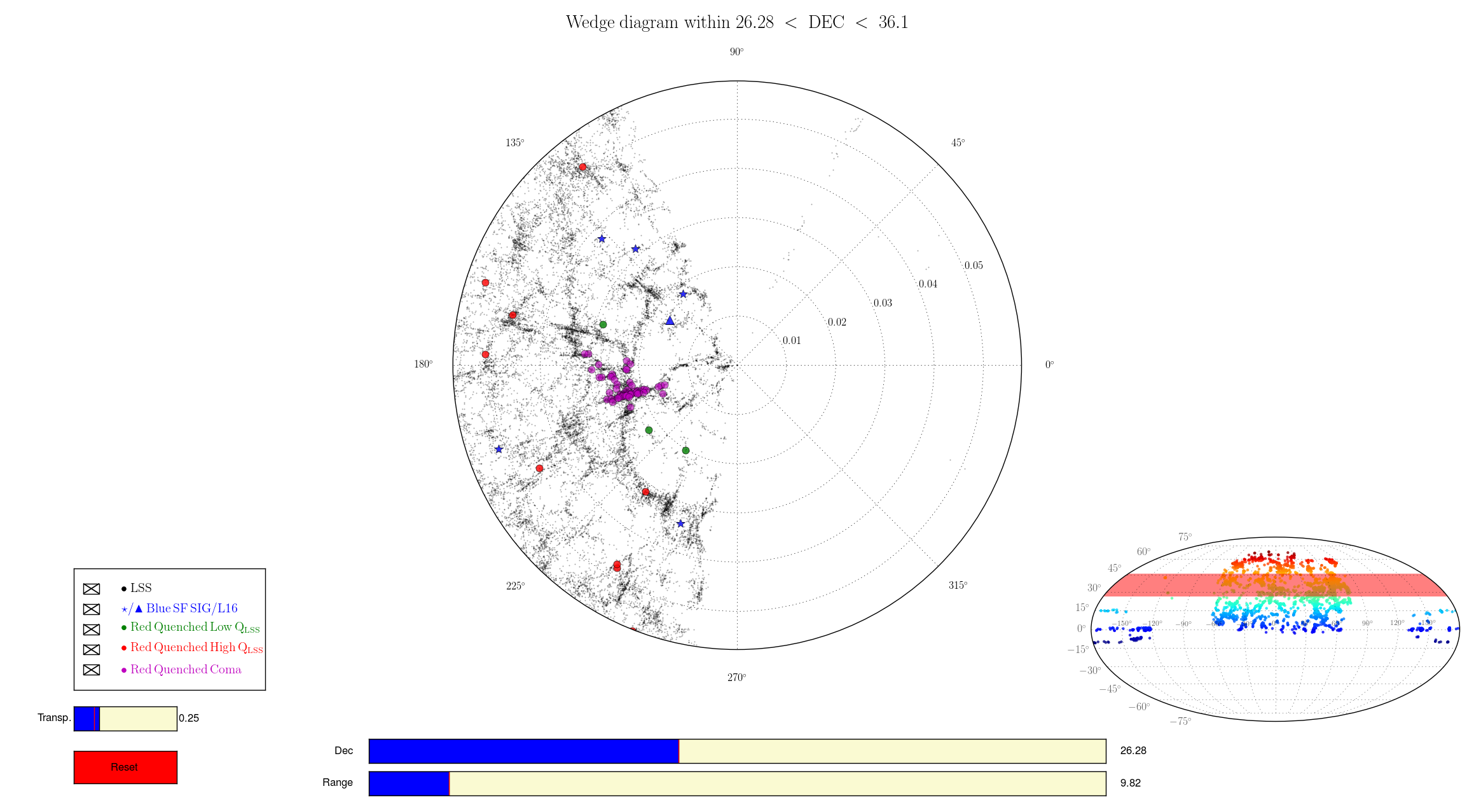}
\caption{Wedge diagram of SDSS galaxies at 26.28 < Declination < 36.1 deg. Blue SF isolated E galaxies from SIG and L16 samples are shown as blue stars and blue triangles, respectively. 
Red, quenched isolated E galaxies in environments of high
(\Q\ $\ge$ -4.75) and low (\Q\ $\le$ -7.0) density are shown in red and green circles, respectively.
As a comparison, E galaxies of the Coma supercluster from L16 are shown in magenta circles. 
}
\label{fig:wedge}
\end{sidewaysfigure*}

For this study, we use the Mollweide and wedge diagram projections of LSSGalPy (see Figs.~\ref{fig:mollweide} and \ref{fig:wedge}, respectively). 
We show blue SF E galaxies in the L16 and SIG samples (blue triangles and stars, respectively) and red, quenched isolated E galaxies in the same samples with the highest and the lowest values of the tidal strength parameter (red circles for isolated galaxies with $\rm Q_{LSS}~\geq~-4.75$ and green circles for isolated galaxies with $\rm Q_{LSS}~\leq~-7.0$, respectively). For comparison, we also show the location of E galaxies in the Coma supercluster \citep[magenta circles,][]{Gavazzi+2003,Lacerna+2016} concerning the LSS (depicted by black points). 

We observe that red, quenched isolated ellipticals with high $\rm Q_{LSS}$ are 
closer to dense structures of the LSS, and isolated ellipticals with low $\rm Q_{LSS}$ are predominantly distributed in lower density regions, in agreement with \citet{Argudo-Fernandez+2016}. Not all the blue SF isolated elliptical galaxies present the lowest values of the $\rm Q_{LSS}$, in other words, they are not the most isolated ellipticals within the samples. Therefore, we observe that they are mainly located in low- and intermediate-density regions but not around dense structures. This supports our results where blue SF isolated ellipticals are mainly located in low-density environments, but 
if an isolated elliptical galaxy is in a large-scale low-density environment, it is not a necessary condition to be a blue SF galaxy. 

%%%Fig

\end{appendix}
\end{document}